\documentclass[preprint]{aastex}
\usepackage{emulateapj5}
\def\stacksymbols #1#2#3#4{\def\theguybelow{#2}
        \def\verticalposition{\lower#3pt}
        \def\spacingwithinsymbol{\baselineskip0pt\lineskip#4pt}
        \mathrel{\mathpalette\intermediary#1}}
\def\intermediary #1#2{\verticalposition\vbox{\spacingwithinsymbol
        \everycr={}\tabskip0pt
        \halign{$\mathsurround0pt#1\hfil##\hfil$\crcr#2\crcr
                \theguybelow\crcr}}}
\def\lta{\stacksymbols{<}{\sim}{2.5}{.2}}
\def\gta{\stacksymbols{>}{\sim}{3}{.5}}

\begin{document}

\title{ABSENCE OF DWARF GALAXIES AT HIGH REDSHIFTS: 
EVIDENCE FROM A GALAXY GROUP}

\author{William G. Mathews\altaffilmark{1}, 
Laura Chomiuk\altaffilmark{1},
Fabrizio Brighenti\altaffilmark{1}\altaffilmark{2},
\& David A. Buote\altaffilmark{3}}
\altaffiltext{1}{University of California Observatories/Lick Observatory,
Department of Astronomy and Astrophysics,
University of California, Santa Cruz, CA 95064, 
mathews@ucolick.org}

\altaffiltext{2}{Dipartimento di Astronomia,
Universit\`a di Bologna,
via Ranzani 1,
Bologna 40127, Italy, 
brighenti@bo.astro.it}

\altaffiltext{3}{University of California at Irvine, 
Dept. of Physics \& Astronomy, 4129 Frederick Reines Hall,
Irvine, CA 92697, buote@uci.edu}

\vskip .2in

\begin{abstract}
The galaxy group NGC 5044 consists of a luminous giant 
elliptical galaxy surrounded by a cluster of 
$\sim 160$ low luminosity and dwarf galaxies, 
mostly of early type.
The cumulative
projected radial distribution
of dwarf galaxies in the NGC 5044 group,
unlike distributions of more luminous galaxies 
in rich clusters, does not
follow a projected dark matter (NFW) profile. 
A deficiency or absence
of low luminosity galaxies is apparent in NGC 5044 within
about 350 kpc.
Most of the dwarf galaxies in NGC 5044
entered the virial radius at redshifts
$z \lta 2-3(a_f/0.25)$,
where $a_f = 1/(1 + z_f)$ is the epoch of
group formation, and very few entered
during redshifts $z \gta 2-3(a_f/0.25)$.
The peculiar, non-NFW 
shape of the projected cumulative dwarf galaxy
distribution in NGC 5044 within 350 kpc 
resembles the characteristic
cumulative distribution of
dark subhalos that are also known to be relatively young.
Dynamical friction is unlikely to explain the
apparent lack of group member galaxies at small
radii in NGC 5044.
\end{abstract}



\section{Introduction}

The early evolution of dwarf galaxies is of interest 
because of the possible great antiquity 
of these galaxies and 
because they lie near the lowest 
mass scale for hierarchical assembly
in which low mass galaxies merge to form normal galaxies  
of larger mass.
Efforts to understand the evolution of these faint galaxies 
have been frustrated because of the difficulty 
of direct observation at cosmologically relevant redshifts. 
Theoretical models of the formation of dwarf galaxies 
are complicated by the poorly understood 
role of supernova feedback and cosmic photoionization that 
may limit star formation at certain redshifts.
Finally, there is much consternation about 
N-body calculations that predict many more dwarf 
galaxies than are actually observed 
(Moore et al. 1999).

We describe here a simple means of determining the 
space density of dwarf galaxies in the early universe 
by observing their projected spatial distribution 
in galaxy groups.
Of particular interest are galaxy groups, such as 
NGC 5044, that contain large numbers of 
dwarf galaxies and for which the virial mass is 
well known from X-ray observations.
These groups are often dominated by a single, 
centrally-located luminous elliptical galaxy.
Since the masses of dwarf galaxies are very small 
compared to the group virial mass, their dynamics are 
very similar to dark matter particles.
As dark halos grow by accretion from the inside out, 
so does the spatial 
distribution of associated dwarf galaxies. 
Dwarf galaxies 
currently orbiting closest to the central elliptical
galaxy must be among the oldest galaxies in the group.

It is therefore of considerable interest that the cumulative 
projected radial 
distribution of dwarf galaxies in the NGC 5044 
group does not follow the 
standard projected mass distribution 
of the dark matter halo.
Instead, the number of dwarf galaxies per unit dark matter 
appears to be lower within a projected radius 
$R \approx 350$ kpc. 
We show here that this central shortfall implies that 
dwarf galaxies were underabundant relative to the ambient 
dark matter at redshifts $z \gta 3$. 
This cosmic variation of dwarf galaxy density 
is opposite to that recently proposed 
to explain the increasing number of 
dwarf galaxies in richer clusters.

\section{Observed Galaxies in NGC 5044}

On first inspection, 
the central elliptical galaxy in 
the NGC 5044 group, with luminosity 
$L_B = 4.5  \times 10^{10}$ $L_{B,\odot}$, completely dominates 
the optical image of the cluster.
A closer look reveals a cluster of 
about 160 low-luminosity galaxies
surrounding the central elliptical that extends 
out to 500 kpc and beyond
(Ferguson \& Sandage 1990; FS).
At 33 Mpc (e.g. Tonry et al. 2001) 
the absolute magnitudes of the non-central galaxies range 
from $M_B = -19$ to $-13$, i.e.
$ 2 \times 10^7 \lta L_B \lta 9 \times 10^9$ $L_{B,\odot}$.
About $80$ percent of the non-central group members 
appear to be spheroidal or dE galaxies (FS). 
Most or all of these galaxies are dwarfs, 
depending on one's threshold for dwarf designation. 
Unfortunately, cluster membership has not yet been
fully confirmed with radial velocity measurements, 
but we have removed
several Ferguson-Sandage galaxies from the
membership list -- FS68,
FS102, and FS137 -- because their velocities differ
by more than $\pm 800$ km s$^{-1}$ from that of the
central elliptical (NED, Garcia 1993).
Nevertheless,
the remarkably high space density of these galaxies and their 
symmetric distribution largely 
confirm their physical association with the 
central elliptical (NGC 5044).
In any case, we assume here that most of the 
galaxies regarded as members or likely members 
by Ferguson \& Sandage are indeed members.

Of particular interest is the radial distribution 
of the galaxies in NGC 5044. 
Carlberg, Yee \& Ellingson (1997) and 
Biviano \& Girardi (2003)
have shown that the mean surface number density 
of galaxies in richer clusters closely follows the NFW profile 
expected for the underlying dark matter 
(Navarro, Frenk, \& White 1997).
The luminosities of the galaxies considered 
by Carlberg et al. are much larger 
than those found in NGC 5044. 
An NFW distribution for optical galaxies would be expected 
if galaxies entered the virial mass of the 
cluster evenly over time
and suffered no significant radial migration 
by dynamical friction not also experienced by the dark matter.

Remarkably, the projected distribution of 
the low luminosity galaxies in 
NGC 5044 differs dramatically from an NFW profile.
Instead of the NFW central peak, 
the satellite galaxies in NGC 5044 have 
a King-like core that extends to $\sim 150$ kpc 
(Ferguson \& Sandage 1990).
This difference is apparent in Figure 1 
where we compare the cumulative number of non-central galaxies 
as a function of projected radius $R$ with an NFW 
mass surface density profile
(Bartelmann 1996). 
The current virial mass and radius of NGC 5044,
$M_{v,o} \approx 4 \times 10^{13}$ $M_{\odot}$ 
and $r_{v,o} = 887$ kpc, 
are known from X-ray observations
(Buote, Brighenti \& Mathews 2004).
In Figure 1 both distributions are normalized to 
agree at $R = 350$ kpc.
A Kolmogorov-Smirnov test applied within 350 kpc 
reveals that the observed cumulative 
galaxy distribution in NGC 5044 has a nearly vanishing likelihood 
of being consistent with the NFW profile. 
The normalization that we have chosen, forcing agreement 
at $R = 350$ kpc, is reasonable because the two 
distributions then agree reasonably well in the range
$325 \lta R \lta 500$ kpc. 
For $R \gta 500$ kpc the number of identified galaxies 
falls below the NFW profile.
We assume that this discrepancy is due to an incompleteness in 
identifying cluster members, but the projected 
radius where the incompleteness sets in could be somewhat 
smaller.

Galaxies that joined the NGC 5044 
group at high redshift when the 
virial radius was smaller have orbits with maximum 
radial excursions near this virial radius. 
These galaxies spend most of their time 
orbiting near this same radius.
As we show below, dynamical friction is relatively 
ineffective for dwarf galaxies. 
Therefore, the numerical shortfall 
of low luminosity galaxies in the 
core of NGC 5044 ($R \lta 350$ kpc) 
suggests that proportionally fewer 
galaxies entered the virial radius at high redshift. 
In the following section we develop a simple dynamical 
model for the evolution of orbiting galaxies in NGC 5044
that can be used to quantify the relative absence 
of low-luminosity galaxies in the early universe.

The shallow projected distribution of galaxies around NGC 5044 
may not be unique.
Klypin et al. (1999) note the lack of dwarfs in
low density regions like the Local Group.
Smith, Martinez \& Graham (2003) find that the radial  
surface density of dwarf galaxies 
varies as $R^{-0.6 \pm 0.2}$ around a sample of 
isolated elliptical galaxies;
this is flatter than NFW except for the innermost regions.
Madore, Freedman \& Bothun (2004) also find that the 
surface density of satellite galaxies near 
individual ellipticals 
declines as $R^{-0.5}$, while earlier studies 
by Vader \& Sandage (1991) and Lorrimer et al (1994) 
indicate a steeper decline, $R^{-1}$.
Dwarf galaxies in the Fornax cluster have a shallower 
cumulative radial distribution than 
the giant galaxies (Drinkwater et al. 2001). 
Very recently, Pracy et al. (2004) have discovered a relative 
deficiency of dwarf galaxies in the central regions 
of the cluster Abell 2218. 
The physical scale of the region of dwarf depletion, 
$\sim 400$ kpc, is very similar to that in NGC 5044.
In Abell 2218 
the ratio of the projected numbers of dwarf to giant  
galaxies decreases with increasing surface density of giant 
galaxies, as noted by Phillipps et al. (1998).

\section{Dynamical Model for Satellite Galaxies}

We estimate the expected radial distribution of galaxies 
in NGC 5044 by following the orbits of cluster 
member galaxies in the time-dependent
dark matter potential of the cluster.
The galaxies are regarded as test particles without structure.
Newly accreted galaxies are 
assumed to arrive at the virial radius having 
fallen from the local turnaround radius.
The orbits are not entirely radial since galaxies acquire 
some angular momentum due to attractions 
by neighboring inhomogeneities.
After entering the virial radius, galaxies 
orbit until the present time when their projected distances 
from the center of the cluster can be compared with
the radial distribution observed in NGC 5044.

The first model we consider is a galaxy cluster 
in which the local space density of galaxies is proportional 
to that of the dark matter, as if a fixed fraction of 
baryons formed into low-luminosity galaxies 
at a very early time. 
In this case the dynamics of dark matter 
and surviving dwarf galaxies are identical. 
Dark matter halos grow from the inside out by radial 
accumulation and accretion by mergers. 
The mass assembly histories of dark halos are universal 
and scale 
invariant, apart from differences in formation times.
From studies of N-body simulations in a $\Lambda$CDM 
cosmology it has been determined that the mass of 
virialized dark matter 
increases with decreasing redshift $z$ 
to its current value $M_{v,o}$ according to
\begin{equation}
M_v = M_{v,o} e^{-2 a_f z}
\end{equation}
where $a_f = 1/(1 + z_f)$ defines the 
halo ``formation'' time (Wechsler et al. 2002). 
We adopt $a_f = 0.25$ here, but the redshifts we derive 
can be easily scaled to other values of $a_f$ by 
keeping $a_f z$ constant.
At every time or redshift the radial profile of dark 
matter is described by an NFW profile 
(Navarro, Frenk \& White 1997) that is 
parameterized by the virial mass $M_v$ and the concentration
\begin{equation}
c(M_v,z) = 450 (M_v/M_{\odot})^{-0.1283} (1 + z)^{-1}
\end{equation}
(Wechsler et al. 2002), 
based on a $\Lambda$CDM cosmology with
$\Omega_o = 0.3$, $\Omega_{\Lambda,o} = 0.7$
and $H_o = 70$ km s$^{-1}$ Mpc$^{-1}$.

For the idealized circumstance 
in which dwarf galaxies observed in clusters today 
all formed at some very early 
time, they enter the virialized halo at the same 
rate as the dark matter. 
With this simple model the 
virialized mass $M_{v,i}$ when the ith galaxy first 
entered the virialized halo can be found from 
\begin{equation}
M_{v,i} = {\cal R}M_{v,o}
\end{equation}
where $0 \le {\cal R} \le 1$ is a random number.
The redshift $z_i$ at entry can be found from Equation (1) 
and the corresponding cosmic time $t_i(z_i)$ is 
determined by the $\Lambda$CDM cosmology. 

The virial radius at time $t_i$ is 
\begin{equation}
r_{v,i} = \left( { 3 M_{v,i} 
\over 4 \pi \Delta_i \rho_{c,i}}\right)^{1/3}
\end{equation}
where $\rho_{c,i} = 3 H_i^2 / 8 \pi G$ is the 
critical density and the appropriate $\Lambda$CDM 
overdensity factor is
\begin{equation}
\Delta_i = 18 \pi^2 + (\Omega(z_i) - 1)
[82 - 39(\Omega(z_i) -1)].
\end{equation}
The velocity of galaxies arriving at $r_{v,i}$,
\begin{equation}
u_{v,i} = (G M_{v,i}/r_{v,i})^{1/2},
\end{equation}
is the freefall velocity from the turnaround 
radius $r_t = 2 r_{v}$, and we ignore the small
dark energy term, i.e. 
$\Lambda r_{v}^2 \ll G M_{v}/r_{v}$.

If the dark matter distribution were perfectly 
smooth, the orbits of dark matter particles and 
galaxies would be perfectly radial.
In this case the dark matter 
halos would be more centrally concentrated than the NFW 
profile (Bertschinger 1985).  
However, because of inhomogeneities in the dark matter 
distribution, random velocities are induced by mutual 
gravitational attraction.
Consequently, galaxies arriving at the virial radius 
have acquired some angular momentum that is essential 
in determining the NFW profile. 
Considering the importance of this effect and the large 
number of available N-body dark matter calculations, it is 
surprising that the angular momentum distribution 
has not been discussed in full detail.
We are fortunate, however, that Tormen (1997) 
and Ghigna et al. (1998) have plotted the distribution 
of circularity $\epsilon = J/J_c$ at $r_v$ where
$J$ is the specific angular momentum of dark matter particles
and $J_c = u_v r_v$ is the maximum angular momentum 
at the virial radius;
$\epsilon = 1$ for circular orbits and 
$\epsilon = 0$ for radial orbits. 

Based on these results we model the probability density 
for circularity $p'(\epsilon) = dp/d\epsilon$ with a third order 
polynomial, subject to the constraints that 
$p'(0) = p'(1) = 0$ and normalized so the 
integral of $p'(\epsilon)$ from 0 to 1 is unity.
The resulting distribution is 
\begin{equation}
p'(\epsilon) = a \epsilon^3 - (1.5 a + 6) \epsilon^2 + (0.5a + 6) \epsilon 
\equiv A \epsilon^3 - B \epsilon^2 + C \epsilon
\end{equation}
where
\begin{equation}
a = 60 (1 - 2 \langle \epsilon \rangle)
\end{equation}
and $\langle \epsilon \rangle$ is the mean circularity.
The results of Tormen suggest that $\langle \epsilon \rangle \approx 0.5$,
for which $p'(\epsilon)$ becomes a symmetric quadratic.
However, the circularity distribution of 
Ghigna et al. is skewed toward larger $\epsilon$ so we also 
entertain this possibility.
Random values of $\epsilon$ can be found from 
\begin{equation}
{\cal R} = \int_0^\epsilon p'(\epsilon) d\epsilon
= (A/4)\epsilon^4 - (B/3)\epsilon^3 + (C/2)\epsilon^2.
\end{equation}
All galaxies have velocity $u_{v,i}$ given by Equation (6) 
when they first enter the virial radius; the angular 
momentum $J = J_c \epsilon$ is used to allocate the 
direction of this velocity.

The galaxy orbits are found by solving
\begin{equation}
{d {\bf r} \over dt} = {\bf u} ~~~{\rm and}~~~
{d {\bf u} \over dt} = - {G M(r) \over r^2}{ {\bf r} \over r}.
\end{equation}
Within the virial radius the NFW mass distribution is 
\begin{equation}
M(r) = M_{v}(t) {f(y) \over f(c)}
\end{equation}
where $f(y) = \ln(1 + y) - y/(1+y)$ and $y = cr/r_v$.
At higher redshifts, when the virial mass grows more rapidly, 
most entering galaxies orbit entirely within the increasing 
virial radius. 
At lower redshifts, however, some galaxies move briefly outside 
$r_v(t)$.
In this case we use the exterior dark matter mass density 
distribution corresponding to the density 
derived by Barkana (2003)
\begin{equation}
\rho = \rho_{v+} (r/r_v)^{-1.3} ~~~r > r_v.
\end{equation}
The density just beyond the virial radius is found from 
\begin{equation}
\rho_{v+} = {1 \over 4 \pi r_v^2 u_v} {d M_v \over dt}.
\end{equation}

Galactic orbits are computed from the time $t_{v,i}$ when they 
first reach the virial radius until the current time
$t_n = 13.5$ Gyrs. 
At time $t_n$ the radius of each galaxy $r_i$ is randomly 
projected on the plane of the sky,
\begin{equation}
R_i = [1 - (2{\cal R} - 1)^2]^{1/2} r_i.
\end{equation}

\section{Results}

The thin lines in Figure 1 show the results of two 
dynamical models calculated as described in the previous section.
For these models, and those discussed below, 
the Monte Carlo simulations are continued until 
1210 galaxies occupy the region $R < 350$ kpc at time $t_n$.
This is ten times the observed number in this region.
The total number of orbits required
to achieve this limit is typically almost twice as large.
The computed projected radial 
distributions are normalized to agree with the NFW mass 
distribution at a projected radius $R = 350$ kpc.
The light curve (actually a histogram) in Figure 1 
labeled $\langle \epsilon \rangle = 0.5$ 
was computed with this constant mean circularity 
for all orbits during the evolution of the dark halo. 
Since Equation (3) requires that 
the arrival of galaxies into the virialized dark halo is 
identical to that of the dark matter, their cumulative 
distribution should be identical to the NFW profile.
The agreement in Figure 1 is seen to be very good indeed, 
particularly considering the simplicity of our model. 
The mean assembly history for the dark halo 
implicitly allows for all past halo-halo mergers.
When two halos merge, orbiting bound galaxies
also merge in a manner that approximately conserves the  
orbital phase space density, and this maintains their 
relative mean locations in the merged dark halo potential.

As the projected radius $R$ approaches the current virial 
radius of NGC 5044, $r_{v,o} = 887$ kpc, the number of 
galaxies in the $\langle \epsilon \rangle = 0.5$ 
calculation in Figure 1 falls about 10 percent below the NFW 
expectation.
This deviation probably occurs because 
there are too many ($\sim 100$) galaxies at 
radii $r > r_{v,o}$ at time $t_n$. 
This discrepancy is insensitive to the value 
of the mean circularity $\langle \epsilon \rangle$, 
i.e. nearly identical results are obtained 
for $\langle \epsilon \rangle = 0.3 $ or 0.7.
However, if $\langle \epsilon \rangle$ is assumed to 
increase steadily with the virialized mass, 
$\langle \epsilon \rangle = M_v/M_{v,o}$, 
the discrepancy is somewhat reduced. 
This result with time-varying $\langle \epsilon \rangle$, 
illustrated in Figure 1 as 
``var $\langle \epsilon \rangle$'', 
has no independent dynamical motivation. 
Other small procedural changes, perhaps in Equation (6), could be 
considered to reduce the discrepancy near the virial radius, 
but we have not explored this since the results shown 
in Figure 1 
are certainly good enough for the region of most 
interest, $R \lta 500$ kpc.

Figure 2 shows the average halo entry redshift 
$\langle z_i \rangle$ for galaxies in annular bins 
at time $t_n$ for the two dynamical models shown in Figure 1
(open and filled circles).
The redshift dispersions are mostly intrinsic, 
and do not simply reflect the finite number of orbiting 
galaxies in each projected bin.
As expected, the mean arrival redshift 
rises sharply toward the center of the cluster. 
Provided the space density of low luminosity galaxies 
is proportional to the dark matter density at all redshifts,
as imposed by Equation (3), the increase in mean redshift
with decreasing projected radius $R$ is robust, 
essentially independent of 
the value or time variation of $\langle \epsilon \rangle$.
In the following we assume $\langle \epsilon \rangle = 0.5$.

The sense of the discrepancy in Figure 1 between 
the cumulative number of galaxies in NGC 5044 
and the NFW profile indicates that the space density of 
low luminosity galaxies relative to dark matter 
was lower at higher redshifts. 
This discrepancy can be incorporated into 
our models if we relax the 
assumption of uniform, early galaxy formation (Equation 3).
We shall consider only models in which the 
space density of dwarf galaxies is constant 
or increases monotonically with time.

Suppose for example that the satellite galaxies in NGC 5044 
began to form at a redshift $z_c$ when the halo mass was 
$M_c$ and increased as a power law afterward. 
In this case the probability density for arrival 
of a galaxy into the virialized halo is 
$p'(M_v) = p_o (M_v - M_c)^q$. 
When $p'(M_v)$ is properly normalized, 
the virial mass at which a random galaxy enters the halo is
\begin{equation}
M_{v,i} = M_c + (M_{v,o} - M_c) {\cal R}^{1/1+q}
\end{equation}
where ${\cal R}$ is a random number. 
This reduces to Equation (3) when $M_c = q = 0$.

Figure 3 shows three evolutionary sequences in which galaxies 
are assumed to begin forming at the earliest possible 
epoch ($M_c = 0$) and increase in number afterward 
according to a power law, $p' \propto M_v^q$. 
Of the three possible powerlaws, $q = 2$ is seen to follow 
best the observed number of galaxies in NGC 5044 at small $R$.
As before, all solutions are normalized at $R = 350$ kpc.
Solutions with $q = 0.75$ ($q = 3.5$) produce too 
many (few) galaxies within 350 kpc, 
but the $q = 2$ histogram follows the observed 
histogram almost perfectly in this region.
Nevertheless, none of these pure powerlaw models
produce the slope change at $\sim 350$ kpc that is 
apparent in the observations. 
If the $q = 2$ solution is correct, 
the observations would have to begin to be incomplete 
just where this slope change occurs. 
This is possible but implausible. 
It seems unlikely that Ferguson \& Sandage 
would have missed identifying $\sim 50$ member galaxies in the 
region $350 \lta R \lta 500$ kpc. 
The total number of galaxies expected inside the projected 
virial radius 887 kpc at $t_n$ is 237, 305 and 352 for 
orbits with $q = 0.75$, 2 and 3.5 respectively.

In Figure 4 we plot a 
better model in which no 
small galaxies form before redshift $z_c = 3.32$ 
($M_c = 0.19 M_{v,o}$), followed by a powerlaw rise with 
$q = 0.1$.
This solution fits the observations almost perfectly
in $R \lta 350$ kpc and continues to increase thereafter 
in approximate proportion to the dark matter. 
The mean binned redshifts for this dynamical model, 
plotted as open squares in Figure 2, shows that the 
observations are consistent with very few galaxies 
in NGC 5044 having accretion redshifts $\gta 2$.

Since the powerlaw rise with $q = 0.1 < 1$ is abrupt,
we also considered a variety of step function models
for $p'(M_v)$. 
In these models $q \rightarrow 0$ 
and the virial mass at the time of galaxy arrival is 
\begin{equation}
M_{v,i} = M_c + {\cal R}(M_{v,o} - M_c)
\end{equation}
where the halo mass $M_c$ 
corresponds to redshift $z_c = -2\ln(M_c/M_{v,o})$.
The step function solution with $z_c = 2.77$ shown 
in Figure 4 also agrees very well with the NGC 5044 
observations, assuming some small incompleteness 
at $R \gta 350$ kpc 
and few foreground or background interlopers.

As a final alternative model we show in Figure 
4 a solution in which the probability density 
$p'(M_v)$ rises initially as a power law 
until $M_v = M_*$ and is constant thereafter. 
For models of this type the mass at which galaxies 
arrive in the halo is given by 
\begin{equation}
M_{v,i} = M_* ({\cal R}/{\cal R}_*)^{1/p+1}
\end{equation}
if 
\begin{equation}
{\cal R} < {\cal R}_* \equiv [ 1 + (p+1)(M_{v,o}-M_*)/M_*]^{-1}
\end{equation}
and 
\begin{equation}
M_{v,i} = M_*[ 1 - (1 - {\cal R}/{\cal R}_*)/(p+1)]
\end{equation}
if ${\cal R} > {\cal R}_*$.
A solution with $p = 2.5$ and $M_* = 0.4M_{v,o}$ 
($z_* = 1.83$) is shown in Figure 4.
This solution also fits the NGC 5044 data remarkably 
well within 350 kpc and corresponds to a modest 
incompleteness in the observed number of galaxies
at larger radii.

Overall, we are struck by the ease at which our 
various dynamical models 
having a relative absence of high redshift dwarf galaxies 
fit the histogram observed by FS within $\sim 350$ kpc.
This excellent agreement 
gives us confidence that the FS data are
reasonably complete in the dense central region
of NGC 5044 and that
the number of dwarf galaxies per unit dark matter
is indeed much lower
for entry redshifts $z \gta 2.5/(a_f/0.25)$ than afterwards.

In addition, we note that 
the cumulative radial projected distributions of dark subhalos 
in N-body calculations 
(e.g. de Lucia, G. et al. 2004;
Gao et al 2004) are also shallower 
than that of the underlying dark halo and have 
histograms very similar 
to the observed cumulative number of dwarfs in NGC 5044.
This is reasonable since subhalos are continuously 
destroyed, so surviving subhalos are 
relatively young (Gao et al. 2004).
However, the evolution of dark subhalos 
and dwarf galaxies differs  
because the subhalos are known to be disrupted 
and dissolved by tidal forces and dynamical friction.

\section{Dynamical Friction}

Can the scarcity of high redshift dE galaxies in the 
NGC 5044 group be understood as a depletion of central 
galaxies by dynamical friction?
The answer is not so obvious since dynamical friction 
increases with the orbiting mass and the initial 
dark mass associated with each dwarf galaxy may 
have been rather large.
However, if the central region of NGC 5044 is depopulated
by dynamical friction, it 
may be repopulated by galaxies 
that decay by dynamical friction from initially larger orbits. 
To address this question we have 
made a series of orbital calculations 
including an additional dynamical friction term 
in the second Equation (10),
\begin{equation}
\left( {d {\bf u} \over dt}\right)_{df}
= - {\bf u} \cdot 4 \pi \ln\Lambda G^2 M \rho u^{-3}
[{\rm erf}(X) - {2 \over \pi^{1/2}} X e^{-X^2}].
\end{equation}
(Chandrasekhar 1943). 
Here $\rho(r,t)$ is the local density of halo dark matter,
$u = |{\bf u}|$ is the galactic velocity,
$X = u/2^{1/2}\sigma$ and 
$\sigma(r,t)$ is the mean velocity dispersion of dark matter 
particles in the NGC 5044 halo.
According to Hoeft, M\"ucket \& Gottl\"ober (2004), 
the dispersion velocity of dark matter is 
approximately isotropic with maximum value 
$\sigma_{m} = 1.25 \pm 0.20 (G M_v/r_v)^{1/2}$ 
and radial distribution
$\sigma(r)/\sigma_{m} = [2 /(\xi^{-0.5} + \xi^{0.6})]^{1/2}$
where $\xi = 2y$ and $y = cr/r_v$. 
We choose $\ln\Lambda = 3$ as suggested by 
Zhang et al. (2002) and others.
Because of the high density
and compact nature of dwarf galaxies,
we approximate the orbiting galaxies with point masses.

In selecting galaxies we assume 
random halo entry redshifts given by Equation (3).
The mass of the ith galaxy is
$M_i = \Upsilon_B L_{B,i}$ where $\Upsilon_B$ is
an adjustable mass to light ratio 
and the galactic luminosity is randomly selected from 
the power law part of the Schechter luminosity 
function $dN/dL_b \propto L_B^{-1.2}$ 
(which is satisfied by NGC 5044),
\begin{equation}
L_{B,i} = [ L_1^{-0.2} - {\cal R}(L_1^{-0.2} - L_2^{-0.2})]^{-5}.
\end{equation}
Here $L_1 = 2 \times 10^7$ $L_{B,\odot}$ and 
$L_2 = 9.2 \times 10^9$ $L_{B,\odot}$ span the range 
of non-central galaxies in NGC 5044.

The ratio of stellar to dark mass in dwarf galaxies
is uncertain since mass to light ratios $\Upsilon_B$ can only
be observed in the central region containing
luminous stars. 
Observations of several dE galaxies by Geha et al. (2002)
indicate $3 \lta \Upsilon_B \lta 6$, but the total value
could be higher, depending on the severity of 
tidal truncation.
Consequently, we computed galactic orbits 
using the friction term (20) 
with values of $\Upsilon_B = 0 - 100$. 
Circularities were determined by Equation (9) assuming  
$\langle \epsilon \rangle = 0.5$ .

For all values of $\Upsilon_B$ considered  
the cumulative number of galaxies at time 
$t_n$ closely follows the 
NFW profile (normalized at $R = 350$ kpc) and resembles 
very closely the 
$\langle \epsilon \rangle = 0.5$ solution in Figure 1. 
For $\Upsilon_B = 4$ only $\sim2$ percent of the dwarf galaxies
merge into the central E galaxy, 
but this increases to 4 or 15 percent if $\Upsilon_B = 10$
or 100 respectively. 
Even with these very large $\Upsilon_B$, 
the loss of galaxies by inward migration and central mergers 
does not appreciably alter their radial distribution at time 
$t_n$.
For all $\Upsilon_B$ considered 
the redshift distribution of surviving galaxies at $t_n$ 
is essentially the same as that shown in Figure 2 
for the $\langle \epsilon \rangle = 0.5$ solution 
without dynamical friction. 

However, when $\Upsilon_B \gta 30$ we find that 
the average luminosity 
$\langle L_B \rangle$ of surviving galaxies decreases. 
Since the drag force is proportional to 
$M d{\bf u}/dt \propto M^2 \propto (\Upsilon_B L_B)^2$,
the onset of dynamical friction is rather abrupt
as the mass or $L_B$ of orbiting galaxies increases.
This may explain why the luminosity distribution of 
low-luminosity galaxies observed in NGC 5044 
cuts off rather abruptly at $L_B \approx L_2$.
In orbital calculations using 
Equation (3) and including dynamical friction, 
the average luminosity of 
surviving satellite galaxies 
$\langle L_B \rangle$ 
drops by 10 or 20 percent if 
$\Upsilon_B = 15$ or 30 respectively.
For $\Upsilon_B = 65$, $\langle L_B \rangle$ is 
only half as large as the average galaxy observed in 
$L_1 < L_B < L_2$ because a larger fraction of 
these more massive galaxies have computed orbits that 
merge at the center under dynamical friction.
We conclude that the mass to light ratios of the 
(possibly tidally truncated) low luminosity galaxies in 
NGC 5044 cannot be much larger than $\sim 30$.
Summarizing, (1) for reasonable dwarf galaxy mass to light 
ratios $\Upsilon_B$ dynamical friction is not very important, and 
(2) the main effect of increasing $\Upsilon_B$ above 
$\sim 30$ is to deplete the more luminous dwarfs 
relative to those observed in NGC 5044.

\section{Discussion and Conclusions}

We emphasize the distinction between the evolution 
of group member galaxies and dark matter subhalos, 
which has been a source of confusion in recent discussions. 
Subhalos are formed by dark matter inhomogeneities that are 
accreted into the main halo.
As these subhalos undergo
tidal disruption and dynamical friction, 
they melt into the main cluster halo 
so surviving subhalos at any time are comparatively young. 
However, the dense baryonic galactic cores in 
subhalos are very deeply bound and are 
much more resistant to dynamical disruption than 
the surrounding subhalos.
The dynamical distinction between ephemeral subhalos 
and long-lived galactic cores has been discussed 
for massive clusters
by Gao et al. (2004) who use N-body and semi-analytic 
methods to reproduce both the shallow, non-NFW 
cumulative distribution of dark subhalos and the 
nearly NFW distribution of surviving galaxies 
(Carlberg, Yee \& Ellingson 1997).
Within the resolution limit of the calculation 
of Gao et al., 
the surviving galaxies typically have 
lost their extended dark 
halos by tidal disruption and this is likely to be true in 
NGC 5044 as well.

We also emphasize that the final galactic distributions 
calculated here 
refer only to cluster member galaxies that survive today.
Many more massive galaxies that did not survive 
(and are therefore not included in most orbits considered 
here) merged 
to form the giant elliptical at the center of NGC 5044.
An integral over a Schechter luminosity distribution from 
the most luminous orbiting galaxy in NGC 5044 
$L_B = L_2$ to infinite $L_B$ is 
comparable to the luminosity of the central galaxy 
(Jones, Ponman \& Forbes 2000).
This is consistent with the notion that
the central elliptical was
formed by mergers of all first generation galaxies 
with $L_B > L_2$.

The relative absence of dwarf galaxies that entered 
the virial radius of NGC 5044 at high redshifts 
is exactly opposite to the behavior that has been 
proposed to explain the excess of dwarf galaxies 
in rich clusters 
(Klypin et al. 1999;
Bullock, Kravsov \& Weinberg 2000;
Trentham, Tully \& Verheijen 2001;
Tully et al. 2002;
Tassis et al. 2003).
Tully et al. (2002) argue, for example, 
that the relative lack of dwarf 
galaxies in the Ursa Major cluster 
($M_v = 4 \times 10^{13}$ $M_{\odot}$) relative to the 
Virgo cluster ($M_v = 8 \times 10^{13}$ $M_{\odot}$) 
can be understood if most of the dwarfs in Ursa Major 
formed after cosmic reionization at redshift 
$z_{ion} \gta 6$ when the heated baryonic gas  
was less able to form stars. 
Late forming dwarfs also have lower dark matter 
(and gas) densities, further suppressing star formation.
Conversely, it is argued that the Virgo cluster is much older 
and its (more numerous) dwarfs formed before reionization. 

Evidently, dwarf galaxy formation and 
the faint end of the galaxy luminosity function 
may vary with space and 
time due to environmental influences. 
The relatively small number of surviving dwarf galaxies that 
entered the virial radius of NGC 5044 
during redshifts $6 \gta z \gta 2-3$ 
could be explained if 
stars were unable to form in small galaxies 
until $z \sim 2-3$. 
If the dark matter potential 
in low mass halos is sufficiently shallow, 
heating by early supernova feedback or 
cosmic ionizing radiation may have kept the baryonic 
gas from cooling into stars. 
If no gas accumulated in or near low mass dark halos 
prior to redshift $z \sim 2-3$,
only the dark halos of dwarf galaxies entered the 
virial radius of NGC 5044.
To explain the existence of dwarf galaxies in NGC 5044 
at smaller redshifts, 
it must be assumed that some baryonic gas 
eventually accreted into these halos and formed into 
the stars observed today.
This scenario is possible, but unlikely.
According to Susa \& Umemura (2004), 
star formation in dwarfs with 
viral masses greater than $\sim 10^8$ $M_{\odot}$ 
is unaffected by cosmic UV heating. 
For typical stellar mass to light ratios $\Upsilon_{B,*} \sim 5$,
the {\it stellar} masses of the NGC 5044 dwarfs 
all exceed $\sim 10^8$ $M_{\odot}$ and their original virial masses 
were $\sim 10$ times larger.  
Therefore star formation in the dwarfs of NGC 5044 
may be impaired but not completely shut down by cosmic UV heating.
This is consistent with observations of nearby 
dwarf galaxies, many of them less massive than those in NGC 5044, 
that show no evidence that the star formation rate changed either 
at the epoch of reionization 
or at redshifts of $\sim 2-3$ (Grebel \& Gallagher 2004)

The apparent deficiency of high redshift dwarf galaxies 
in NGC 5044 can also be understood 
if these galaxies retained most of their gas at 
all relevant redshifts, provided very few baryons in these galaxies
formed into stars before $z \sim 2-3$, 
as implied by the slow and sustained star formation in many
well-observed local 
dwarf galaxies (e.g. Grebel \& Gallagher 2004).
Star formation in gas-rich dwarfs could be reduced either 
by proximity to NCC 5044 or by gas removal after 
the dwarf entered the virial radius of NGC 5044.
Suppose for example that 
some disturbing process occurred during 
formation of the central galaxy (NGC 5044) -- such as UV emission, 
outflowing winds, AGN activity etc. -- 
that suppressed star formation in nearby dwarf galaxies. 
A necessary condition for this scenario is that 
the central formation starburst of NGC 5044 
preceded
the formation of most of the stars in the surrounding
system of dwarf galaxies.
As discussed already, 
even the least massive of the dwarfs identified in NGC 5044 
is massive enough to retain ionized gas at $\sim 10^4$ K, 
but gas heated to this temperature 
in gas-rich dwarf galaxies may expand sufficiently 
in the dwarf potential to greatly
increase the efficiency of ram-pressure stripping 
after the galaxies entered the virialized hot gas 
in the NGC 5044 group. 
In this case the stellar component of high redshift dwarfs 
in NGC 5044 would be optially faint. 
One possible difficulty with this scenario is that 
the redshift range in Figure 2 corresponding to the 
reduced numbers of dwarfs, $2 - 3 \lta z \lta 6$, corresponds
to almost 2 Gyrs and it is unlikely that an ionizing 
starburst (or AGN activity) can be maintained for this length of time.
If the gas in dwarf galaxies is cooler than $\sim 10^4$ K, 
it is unclear if it can be ram-stripped by the hot 
gas in NGC 5044.

Alternatively, could two-body interactions 
with larger galaxies, dark galaxies or subhalos 
at early times have scattered the orbits 
of dwarf galaxies originally near the center to larger radii in 
the NGC 5044 group? 
Probably not. 
Normal two-body galactic interactions are strongly masked by the 
large background dark matter density in NGC 5044 that 
dominates the force on each dwarf galaxy. 
For example,
the concentrated baryonic mass of the central massive galaxy 
in this group is overwhelmed by the ambient NFW dark mass beyond a
radius of only $20$ kpc. 
If galaxies are inefficient orbital scatterers, 
perhaps the softer potentials of massive 
dark subhalos within $r \sim 350$ kpc in NGC 5044 
strongly perturbed the orbits of dwarfs near the center.
Unfortunately, subhalo perturbations are 
not likely to be effective since (1) X-ray observations 
and detailed N-body calculations verify 
that the dark matter orbiting in the central region of NGC 5044 
is not selectivly depleted in this way 
and (2) normal galaxies in the corresponding 
central regions of rich clusters are not scattered 
outward by dark subhalos.
This is verified by the galaxy plus 
subhalo simulations of Gao et al. (2004) and by 
direct observation (Carlberg et al. 1997).
The NFW distribution of luminous galaxies in rich clusters can 
be understood if these galaxies formed well in advance 
of the cluster potential.

We appeal to observers to study NGC 5044 and other 
similar groups in more detail. 
Group membership, particularly at large radii, must 
be confirmed with accurate radial velocities. 
Measured orbital velocities must also support the relative 
absence of dwarf galaxies at small radii.
Group memberships should be determined out to the virial 
radius and beyond. 
If our interpretation of the 
incompleteness of the Ferguson-Sandage data
is correct, another 50-100 members in NGC 5044 remain to
be discovered in $400 \lta R \lta 900$ kpc (Figure 1).
If possible, estimates of the stellar ages in dwarfs 
as a function of projected radius could constrain our 
argument for a relative deficiency of old dwarfs 
having accretion redshifts $6 \lta z \lta 2-3$.
An extension of group membership to dwarf galaxies of 
even lower luminosity could also be useful.

Our main conclusions are: 
(1) The cumulative projected radial distribution
of dwarf galaxies in the NGC 5044 group 
does not follow a projected NFW profile, 
unlike observed distributions of more luminous galaxies 
in rich clusters. 
(2) Inside about 350 kpc in NGC 5044 
there is a deficiency or absence
of low luminosity galaxies relative to the NFW profile. 
(3) Most of the dwarf galaxies in 
this inner region of NGC 5044 that would have 
entered the virial radius at redshifts
$z/(a_f/0.25) \gta 2-3$ are largely absent;
most of the surviving dwarf galaxies in NGC 5044
first passed through the virial radius at redshifts 
$z \lta 2-3(a_f/0.25)$,
where $a_f = 1/(1 + z_f)$ is the epoch of
group formation.
(4) The peculiar, non-NFW 
shape of the depleted projected cumulative dwarf galaxy
distribution in NGC 5044 within 350 kpc
resembles the characteristic cumulative distribution of
dark subhalos that are also known to be relatively young.
(5) Dynamical friction is an unlikely explanation for the
apparent lack of group member galaxies at small
radii in NGC 5044.
(6) The mass to light ratio $\Upsilon_B$ of dwarf galaxies in 
NGC 5044 is unlikely to exceed $\sim 30$ since otherwise 
the more luminous observed dwarf galaxies would have been lost by 
dynamical friction.  
(7) Our finding that dE galaxies did not appear until 
moderate redshifts is qualitatively 
consistent with the low mean stellar age of $\sim 5$ Gyrs 
observed in dE galaxies (Geha et al 2002), i. e. many or most 
dwarf galaxies are younger than typical giant elliptical 
galaxies.

\vskip.4in
Studies of the evolution of hot gas in elliptical galaxies
at UC Santa Cruz are supported by
NASA grants NAG 5-8409 \& ATP02-0122-0079 and NSF grants  
AST-9802994 \& AST-0098351 for which we are very grateful.



\clearpage
\begin{figure}
\includegraphics[bb=90 216 522 569,angle=270]
{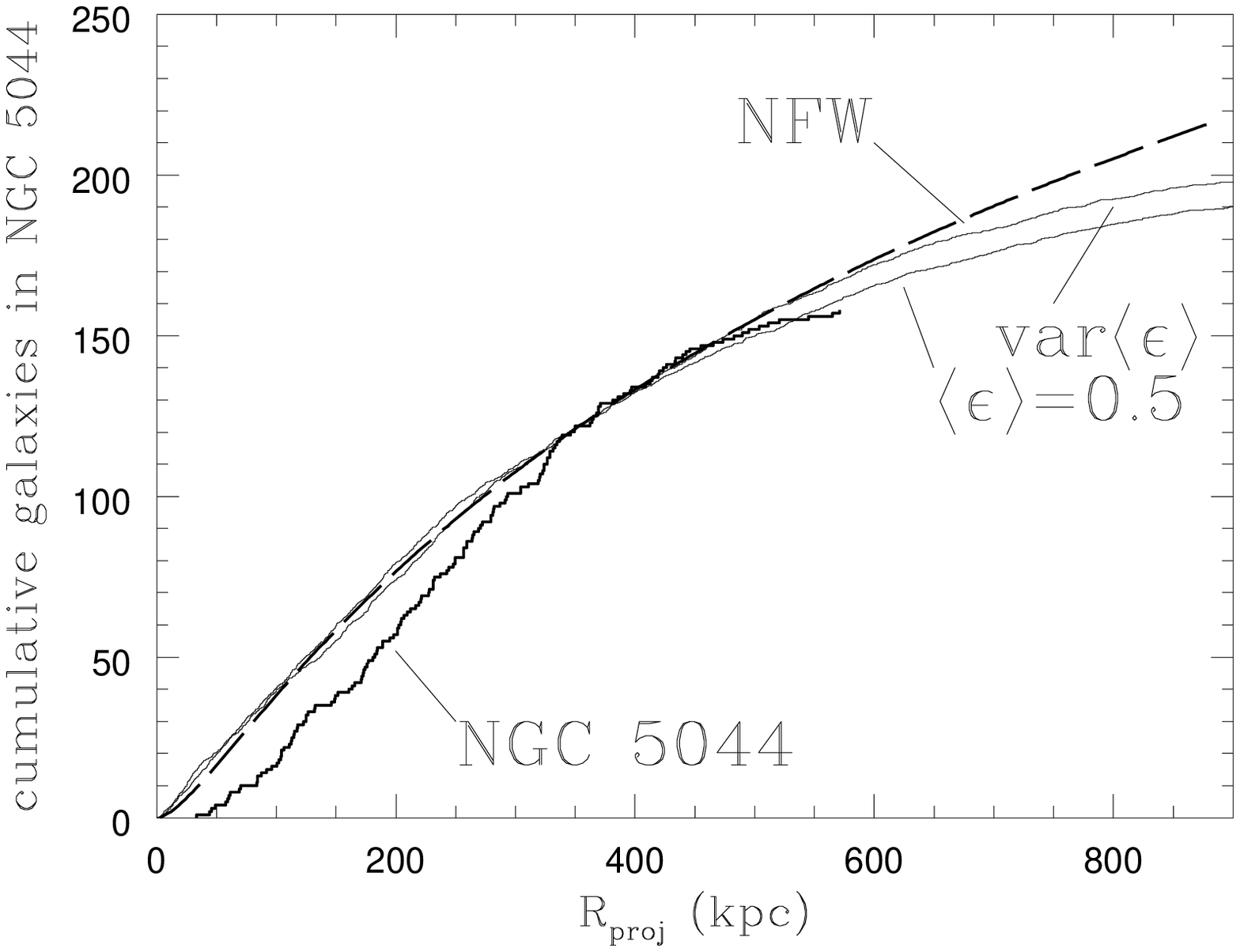}
\vskip.7in
\caption{
The irregular heavy line curve labeled 
``NGC 5044'' is a cumulative histogram of the 
projected radii of 
low luminosity galaxies in the galaxy group NGC 5044.
The heavy dashed line is the cumulative projected 
mass of an NFW profile normalized to agree with the 
NGC 5044 histogram at $R = 350$ kpc.
The two thin lines are dynamical models 
(actually histograms) of the cumulative radial 
distribution of dwarf galaxies in NGC 5044 if the galaxies
entered the virialized halos in proportion to the dark 
matter. The circularity is constant 
for the $\langle \epsilon \rangle = 0.5$ curve and 
increases with the virial mass 
from 0 to 1 in the curve labeled 
``var $\langle \epsilon \rangle$''.
}
\label{fig1}
\end{figure}

\clearpage
\begin{figure}
\includegraphics[bb=90 216 522 569,angle=270]
{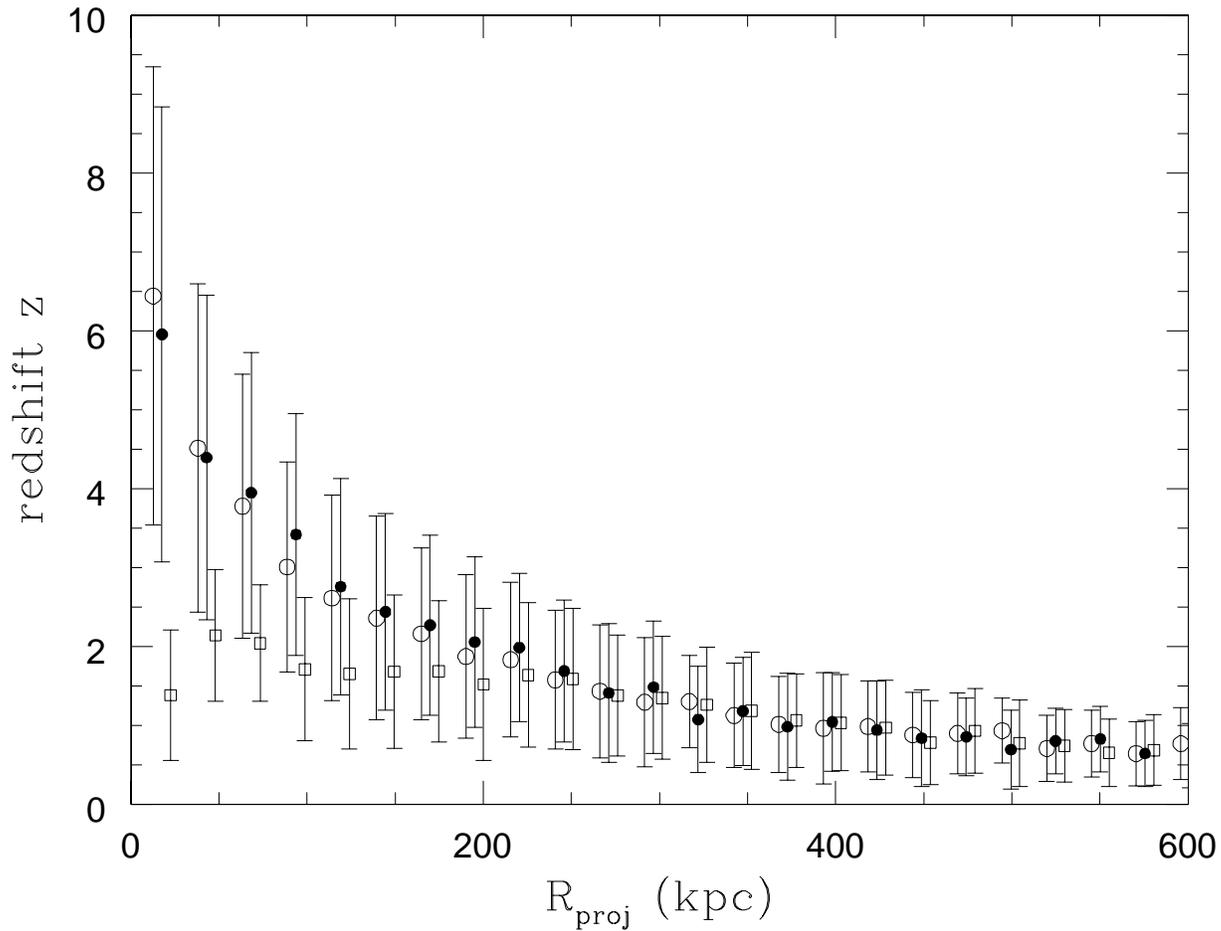}
\vskip.7in
\caption{
The mean redshift and redshift dispersion calculated for 
member galaxies in NGC 5044 in annular bins of projected radius. 
The open circles correspond to the $\langle \epsilon \rangle = 0.5$
model in Figure 1.
The filled circles and open squares correspond respectively to 
redshifts of model galaxies in models 
``var $\langle \epsilon \rangle$'' (Figure 1) 
and ``$z_c = 3.32$, $q = 0.1$'' (Figure 4).
All data points refer to the bins for the open circles but 
the filled circles and open squares 
have been shifted to the right for clarity.
}
\label{fig2}
\end{figure}

\clearpage
\begin{figure}
\includegraphics[bb=90 216 522 569,angle=270]
{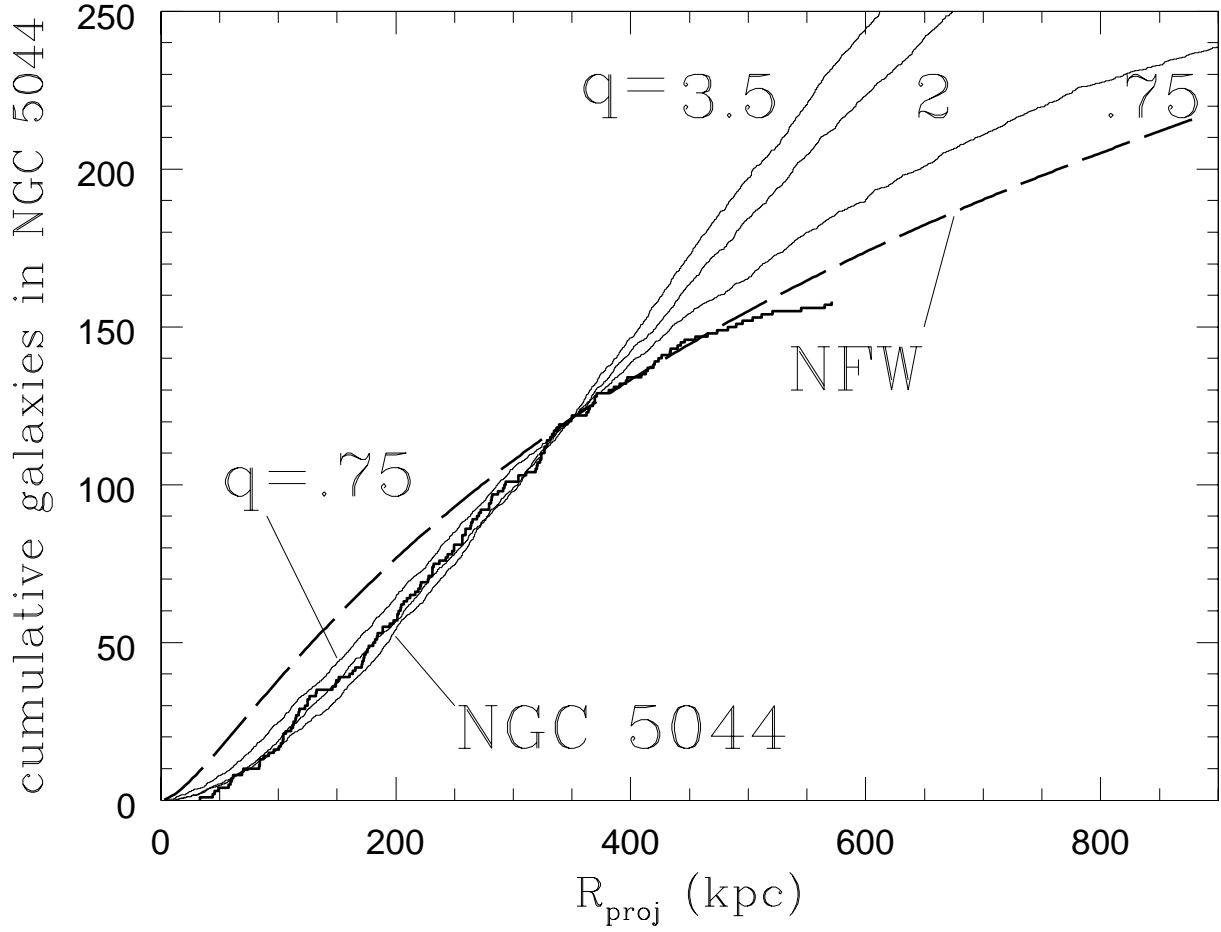}
\vskip.7in
\caption{
The observed histogram for NGC 5044 galaxies and the normalized 
NFW profile are identical to those in Figure 1.
The three light lines are histograms of the cumulative 
number of galaxies in NGC 5044 assuming that they 
entered the virialized halo with probability 
density $p'(M_v) \propto (M_v)^q$ with 
$q = 0.75$, 2 and 3.5. 
}
\label{fig3}
\end{figure}

\clearpage
\begin{figure}
\includegraphics[bb=90 216 522 569,angle=270]
{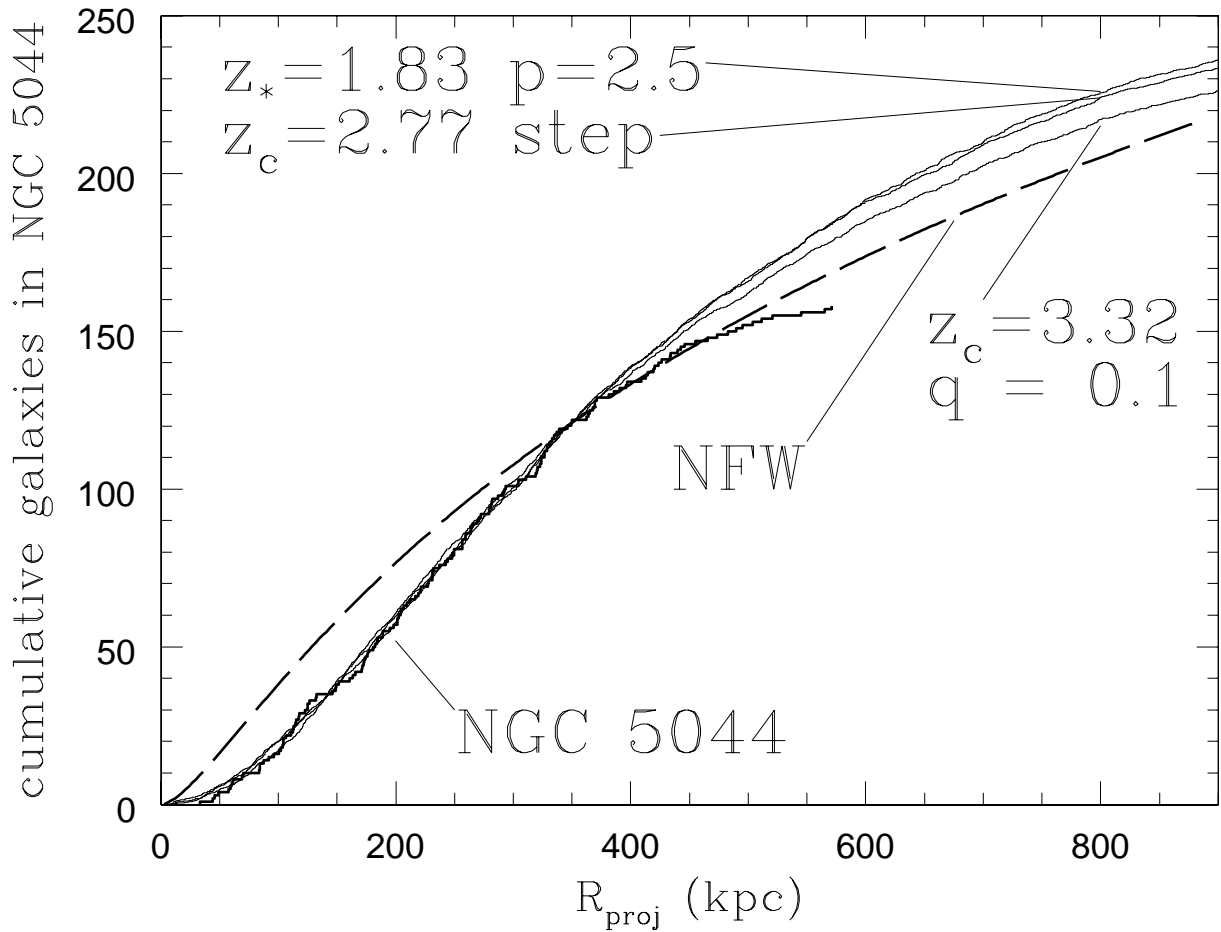}
\vskip.7in
\caption{
Three additional models for the entry and dynamics of 
dwarf galaxies in the NGC 5044 group.
For the histogram labeled ``$z_c = 3.32$ $q = 0.1$'' 
no galaxies entered the halo of the group before redshift 
3.32 when the virial mass was $M_c = 0.19M_{v,o}$, 
but afterwards galaxies entered in proportion to 
$(M_v - M_c)^{0.1}$.
The ``$z_c = 2.77$ step'' curve shows the distribution of galaxies 
resulting from a step function at this redshift.
In the galaxy distribution labeled ``$z_* = 1.83$ 
$p = 2.5$'' galaxies entered the halo of NGC 5044 in 
as $M_v^{2.5}$ until redshift $z_* = 1.83$ and 
in proportion to the dark matter afterward.
The observed histogram for NGC 5044 galaxies and the normalized
NFW profile are identical to those in Figure 1.}
\label{fig4}
\end{figure}

\end{document}